\documentclass[journal, 11pt,one column,twosides]{IEEEtran} 
\usepackage[utf8]{inputenc} 
\usepackage[T1]{fontenc}
\usepackage{url}
\usepackage{ifthen}
\usepackage{empheq} 
\usepackage{cite}
\usepackage{amsmath} 
\usepackage{graphicx}
\usepackage{caption}
\usepackage{subcaption}
\usepackage{color}
\usepackage{epsfig}

\usepackage{amssymb}
\usepackage{amsmath}
\usepackage{amsthm}
\usepackage{latexsym}
\usepackage{setspace}
\usepackage{bbm}
\usepackage{arydshln}
\usepackage{textcomp}
\usepackage{amsfonts}
\usepackage{blindtext}
\usepackage{enumerate}
\usepackage{upgreek}
\usepackage[bottom]{footmisc}
\usepackage{tikz}
\newcommand{\qeed}{\hfill $\blacksquare$}

\usepackage{color}

\theoremstyle{remark}
\newtheorem*{remark*}{Remark}
\newtheorem*{remarks*}{Remarks}

\usepackage{secdot}
\allowdisplaybreaks
\sectiondot{subsection}
\sectiondot{subsubsection}
\usepackage{mleftright}

\usepackage{chngcntr}
\usepackage{fancyhdr}

\begin{document}

\newcommand\MC{{ \ - \!\!\circ\!\! - \ }}

\theoremstyle{theorem}
\newtheorem{theorem}{Theorem}
\newtheorem{corollary}[theorem]{Corollary}
\newtheorem{lemma}[theorem]{Lemma}
\newtheorem{proposition}[theorem]{Proposition}
\theoremstyle{definition}
\newtheorem{definition}{Definition}

\title{Gaussian Data Privacy\\ Under Linear Function Recoverability} 

\author{Ajaykrishnan Nageswaran$^\dag$ }
\maketitle
{\renewcommand{\thefootnote}{}
\footnotetext{
$^\dag$A. Nageswaran is with the Department of
Electrical and Computer Engineering and the Institute for Systems
Research, University of Maryland, College Park, MD 20742, USA.
E-mail: ajayk@umd.edu. This work was supported by the U.S.
National Science Foundation under Grant CCF $1527354$. This paper was presented in
part at the $2022$ International Symposium on Information Theory~\cite{Nages22-2}.
}
}

\maketitle

\maketitle

\begin{abstract}
A user’s data is represented by a Gaussian
random variable. Given a linear function of the data, a querier is
required to recover, with at least a prescribed accuracy level, the function value
based on a query response provided
by the user. The user devises the query response, subject to
the recoverability requirement, so as to maximize privacy of the
data from the querier. Recoverability and privacy are both measured by $\ell_2$-distance
criteria. An exact characterization is provided of maximum user data privacy
under the recoverability condition. 
An explicit optimal achievability scheme for the user is given whose privacy is 
shown to match a converse upper bound.

\end{abstract}

\section{Introduction}
\label{sec:intro}

\label{sec:intro}

A (legitimate) user's data is represented by a Gaussian random variable (rv) and
a querier wishes to compute a given linear function of the 
data from a query response provided by the user. The query response is a suitably
randomized version of the data which the user 
constructs so as to enable the querier to recover  the function value with a 
prescribed accuracy. Under this recoverability requirement, the user 
wishes to maximize privacy of the data from the querier. Both recoverability and
privacy are measured by $\ell_2$-distance criteria.

The contributions of this paper are as follows. In our formulation, the user-provided 
query response should be such that its expected $\ell_2$-distance from the function value
is no greater than $\rho$, 
$\rho\geq 0$. Under this recoverability constraint, we consider a notion of privacy measured by 
the expected $\ell_2$-distance between the user data and the querier’s best estimate of it based on 
the query response, i.e., the corresponding minimum mean-square estimation error (MMSE). We provide 
an exact characterization 
of maximum privacy under the $\rho$-recoverablity requirement as a function of $\rho$,
and specify an explicit query response that attains it. This maximum privacy is shown to be a 
nondecreasing piecewise affine function of $\rho$, and depends on the linear mapping only through 
its rank and singular values. The user implements the optimal query
response by attenuating the function value and adding to it a
suitable independent Gaussian noise. This query response constitutes a multidimensional extension of 
a scheme in~\cite{Wu12} in the separate context of maximizing the MMSE in estimating a 
one-dimensional Gaussian rv on the basis of a one-dimensional randomized version of it under a 
constraint on the expected $\ell_2$-distance between the input and the randomized output, 
where the maximization is over all possible randomization mechanisms satisfying the constraint.

This work is motivated by applications involving analog user data in a database which then must release functional information to a querying consumer. An important goal is to preserve user data privacy while ensuring high accuracy of the released information. For example, when analog biometric data, such as fingerprint or voice patterns, are collected and certain attributes are released, an objective is to safeguard the privacy of individual user data while ensuring high utility of the released information.

Our approach is in the spirit of prior works~\cite{Rebollo10},~\cite{Makhdoumi13},~\cite{Calmon17},~\cite{Liao18},~\cite{Nages19-1},~\cite{Nages19-2},~\cite{Nages22} that deal with maximizing data privacy for a given level of utility of a randomized function of the data. Specifically, for example, in~\cite{Rebollo10},~\cite{Makhdoumi13},~\cite{Calmon17}, a setting where a user possesses private finite-valued data with associated nonprivate correlated data is considered. A randomized version of the nonprivate data is released publicly. The utility constraint is that the expected distortion between the nonprivate and public data should be no more than a specified level. The public data is designed by the user in such a way that privacy measured in terms of the mutual information between the private and public data is maximized. Next, in~\cite{Nages19-1},~\cite{Nages19-2}, user data is represented by a finite-valued rv and a querier which wishes to compute a given function of user data. Privacy is gauged in terms of probability of error incurred by the querier in estimating the user data based on a query response that is a user-provided randomized version of the data. Privacy is maximized under a constraint on the utility of the query response as measured by a conditional probability of error criterion. In all these works, the data that is considered is of the digital type. Our work is motivated by applications involving analog data. This work is a preliminary 
Gaussian counterpart of our earlier works on data privacy under function recoverability for a 
finite-valued rv~\cite{Nages19-1},~\cite{Nages19-2}; notions of privacy and recoverability are 
different, as is the technical approach.

Related works in~\cite{Assodeh16-1,Assodeh16-2} deal with private and nonprivate correlated data
with a given joint distribution. In~\cite{Assodeh16-1}, a Gaussian noise independent of the private and 
nonprivate data is added to the latter and released publicly. The parameters of the Gaussian noise are 
obtained as a result of minimizing the MMSE in estimating the nonprivate data 
from the public data. This minimization is done under a constraint on the MMSE in estimating
the private data from the public data. In~\cite{Assodeh16-2}, first a Gaussian noise independent
of the private and nonprivate data is 
added to the latter, and the sum is quantized and released publicly. In this case, the parameters of the 
Gaussian noise are obtained as a result of maximizing the mutual information between the public
and nonprivate 
data under a constraint on the mutual information between the public and private data. In another related work in~\cite{Diaz21}, motivated by MMSE as a measure of information leakage, a neural network-based estimator of MMSE is characterized. These works involve 
maximizing recoverability under a privacy constraint. In contrast, our goal is to 
maximize privacy under a recoverability constraint.

An important movement in data privacy that has dominated attention over the years is 
differential privacy, introduced in~\cite{Dwork06},~\cite{DworkSmith06} and explored further
in~\cite{McSherry07},~\cite{Bassily13},~\cite{Kasi14},~\cite{Mironov17}, among others. Consider a data vector that represents multiple users’ data. The notion of differential privacy requires that altering a data vector slightly leads only to a near-indistinguishable change in the corresponding probability distribution of the output of the data release mechanism, which is a randomized function of the data vector. A large body of work exists that seeks to maximize
function recoverability under a differential privacy constraint, by minimizing a discrepancy cost 
between function value and randomized output; cf. e.g.,~\cite{Hardt10},~\cite{Geng16},~\cite{Geng20}. Our
alternative approach, i.e., maximizing privacy under a recoverability constraint, can be viewed as a complement to this body of work.

Our model for $\rho$-recoverable linear function computation with associated privacy is 
described in Section~\ref{sec:prelim} and the derivation of $\rho$-privacy is
given in Section~\ref{sec:rho_privacy}. A
closing discussion is contained in Section~\ref{sec:discussion}.

\section{Preliminaries}
\label{sec:prelim}

Let a user's data be represented by a $\mathbb{R}^n$-valued redundant
Gaussian rv $X\sim\mathcal{N}\left(\mathbf{0},I_n\right)$ with covariance matrix
$I_n, \ n\geq 1,$ the identity matrix of size $n$. 
A querier -- who does not know $X$ -- wishes to compute a given linear function of
the user data $AX$, where $A\in\mathbb{R}^{m\times n}$ and has 
rank $1\leq r\leq \min\left\{m,n\right\}$. The querier obtains from the user a
\textit{query response} (QR) $Z$ that is a $\mathbb{R}^m$-valued rv generated by a
conditional distribution $P_{Z|X}$. A QR $Z$ must satisfy the following recoverability condition. 
\vspace{0.05cm}

\begin{definition}
\label{def:rho-recov}
Given $\rho\geq 0,$ a QR $Z$ is 
$\rho$\textit{-recoverable} if 
\begin{equation}
\label{eq:recov1}
\mathbb{E}\left[{\left\Vert AX-Z\right\Vert}^2\right]\leq 
\rho,
\end{equation} 

\noindent where the expectation is with respect to the distribution of $(X,Z)$. 
Such a $\rho$-recoverable $Z$ will be termed $\rho$-QR. We note that the distribution of $Z$ will depend, in general, on $\rho$. 
\end{definition}
\vspace{0.05cm}

\begin{definition}
Given $\rho\geq 0$, the \textit{privacy} of a $\rho$-QR $Z$ satisfying~\eqref{eq:recov1} is
\begin{equation}
\label{eq:def_priv}
\pi_\rho(Z)\triangleq\inf_{g} \ \mathbb{E}\left[{\left\Vert X-g\left(Z\right)
\right\Vert}^2\right] = \text{mmse}\left(X|Z\right)
\end{equation}

\noindent where the infimum is taken over all (Borel) measurable
estimators $g:\mathbb{R}^m\rightarrow\mathbb{R}^n$ of $X$ on the basis of $Z$.
Clearly, the infimum in~\eqref{eq:def_priv} is attained by an MMSE estimator so
that $\pi_{\rho}(Z)$ is mmse$(X|Z)$, where mmse$(X|Z)$ denotes the MMSE in estimating $X$ on the basis of $Z$.
\end{definition}
\vspace{0.05cm}

\begin{definition}
Given $\rho\geq 0$, the maximum privacy that can be attained by a $\rho$-QR $Z$ 
is termed $\rho$\textit{-privacy} and denoted by $\pi\left(\rho\right),$ i.e.,
\begin{equation}
\label{eq:gauss}
\pi\left(\rho\right)\triangleq \sup_{P_{Z|X} : 
\mathbb{E}\left[\left\Vert AX-Z\right\Vert^2\right]\leq\rho}  \pi_\rho\left(Z\right).
\end{equation}
\end{definition}

Our objective is to characterize $\rho$-privacy $\pi(\rho), \ \rho\geq 0 $,
and identify a $\rho$-QR that achieves it. This objective is addressed in Section~\ref{sec:rho_privacy}.

\section{$\rho$-Privacy}
\label{sec:rho_privacy}

Theorem~\ref{thm:gauss_privacy} below provides an exact characterization of $\rho$-privacy in~\eqref{eq:gauss}.
This is done by obtaining
first an upper bound (converse) for $\pi(\rho), \ \rho\geq 0$, and then identifying 
a $\rho$-QR whose privacy meets the bound (achievability).

Throughout the rest of the paper, we consider a particular singular value decomposition
of\footnotemark\footnotetext{Some notation relevant for the rest of the paper: For a
matrix $B,$ we denote its transpose and trace by $B^T$ and tr($B$), respectively.
For a rv $Y$, var$(Y)$ will denote the trace of the covariance matrix of $Y$.}
\begin{equation}
\label{eq:svd}
A=USV^T,
\end{equation} 
\noindent where $U$ and $V$ are, respectively, $m\times m$- and $n\times n$-orthonormal 
matrices. In\footnotemark\footnotetext{The columns of $U$ and columns of
$V$ are the left-singular vectors and right-singular vectors of $A$, respectively.}~\eqref{eq:svd},
the $m\times n$ matrix $S$ containing the singular values of $A$ and represented by 
\begin{equation}
\label{eq:S_matrix}
    S_{ij}=\begin{cases}
    s_k, \ \ \ &i=j=k, \ \ k=1,\ldots,r\\
    0, \ \ \ &\text{otherwise,} \ \ \ \ \ \ \ \ \ \ \ \ \ \ \
      i\in \{1,\ldots,m\}, \  j \in \{1,\ldots,n\},
\end{cases}
\end{equation}

\noindent is such that 
\begin{equation}
    \label{eq:sing_val_asc}
    0< s_1\leq\cdots \leq s_r,
\end{equation}
where $s_1,\ldots,s_r$ are the nonzero singular values of $A$. Let $\tilde{S}$ be the 
$r\times n$-matrix consisting of the first $r$ rows of $S$; the remaining $m-r$ rows of $S$ are all-zero rows. 

Recalling that $U$ and $V$ are orthonormal matrices and using~\eqref{eq:svd},~\eqref{eq:S_matrix}, standard calculations, repeated in Appendix~\ref{app:gauss_standard_calc} for the sake of completeness, show that
$
\text{var}(AX)=\text{tr}\left(AA^T\right)=\sum\limits_{i=1}^{r}s_i^2$ and
\begin{equation}
\label{eq:mmse_XAX}
\text{mmse}\left(X|AX\right)=n-r. \end{equation}

\begin{theorem}
\label{thm:gauss_privacy}
$\rho$-privacy equals 
\begin{equation}
\label{eq:thm_eqn}
\pi(\rho) = n-r+\min\left\{\frac{\rho}{s_1^2},1+\frac{\rho-s_1^2}{s_2^2},\ldots,r-1
+\frac{\rho-\sum\limits_{i=1}^{r-1}s_i^2}{s_r^2},r\right\}, \ \ \ \ \ \   \rho\geq 0.
\end{equation}
\end{theorem}

\begin{figure}[h]
\centering
\includegraphics[scale=0.6]{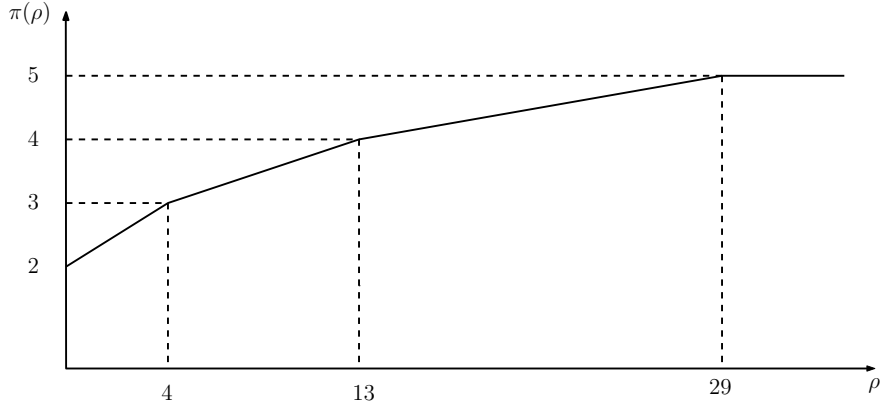}
\caption{$\pi(\rho)$ vs. $\rho.$}
\label{fig:privacy_graph}
\end{figure}

\noindent\textit{Remarks}:  
\begin{enumerate}[(i)]
\item By Theorem~\ref{thm:gauss_privacy} and~\eqref{eq:sing_val_asc},
\vspace{-0.015cm}
\begin{equation}
\label{eq:gauss-priv-alt-form}
\pi(\rho)=
\begin{cases}
n-r+\frac{\rho}{s_1^2}, \ \ \ &0\leq\rho\leq s_1^2\\
n-r+1+\frac{\rho-s_1^2}{s_2^2}, \ \ \ &s_1^2\leq\rho\leq s_1^2+s_2^2 \\\vdots&\vdots
\\
n-r+r-1+\frac{\rho-\sum\limits_{i=1}^{r-1}s_i^2}{s_r^2},\ \ 
\ &\sum\limits_{i=1}^{r-1}s_i^2\leq\rho\leq\sum\limits_{i=1}^r s_i^2\\
n,\ \ \  &\rho\geq\sum\limits_{i=1}^r s_i^2,
\end{cases}\end{equation}
\noindent is piecewise affine in $\rho$. For example, a plot of $\pi(\rho)$ vs.
$\rho$ is given in Fig.~\ref{fig:privacy_graph} for $n=5$, $m=r=3, \ s_1=2, \ s_2=3 $ 
and $s_3=4$.
\item In particular, for $\rho=0$, $\pi(\rho)=n-r$ which, from~\eqref{eq:mmse_XAX}, 
is the error of an MMSE estimator of $X$ on the basis of $AX$. For $\rho\geq 
\sum\limits_{i=1}^r s_i^2=\text{tr}\left(AA^T\right)=\text{var}(AX)$,
$\pi(\rho)=n=\text{var}(X)$ is the error of a MMSE estimator of $X$ 
without any observation.
\item $\rho$-privacy $\pi(\rho)$, $\rho\geq 0$, is a nonincreasing function of each individual 
singular value when the remaining $r-1$ singular values are fixed.
\end{enumerate}
\vspace{0.1cm}

\vspace{0.1cm}

The following Lemmas~\ref{lem:A-sing} and~\ref{lem:dpi} are pertinent to the proof of 
Theorem~\ref{thm:gauss_privacy}. Their proofs are relegated to Appendix~\ref{app:gauss_priv}.

\begin{lemma}
\label{lem:A-sing}
For $\rho\geq 0,$
\begin{equation}
\label{eq:A-sing}
    \sup_{P_{Z|X} : \mathbb{E}\left[\left\Vert AX-Z\right\Vert^2\right]\leq\rho} \ 
    \inf_{g} \ \mathbb{E}\left[{\left\Vert X-g(Z)
\right\Vert}^2\right]= \sup_{P_{\bar{Z}|X} : 
\mathbb{E}\left[\left\Vert SV^{T}X-\bar{Z}\right\Vert^2\right]\leq\rho} \inf_{g} \ 
\mathbb{E}\left[{\left\Vert X-g(\bar{Z})
\right\Vert}^2\right],
\end{equation} 
\noindent where the $\mathbb{R}^m$-valued rv $\bar{Z}$ represents a generic stand-in for a $\rho$-QR under recoverability of $SV^{T}X$.
\end{lemma}
\vspace{0.1cm}
\noindent\textit{Remark}: Since $V$ is orthonormal and $X\sim\mathcal{N}\left(\mathbf{0},I_n\right)$, 
$\bar{X}=V^TX$ has the same distribution as $X$. Hence, the right-side of~\eqref{eq:A-sing} can be 
interpreted as $\rho$-privacy under recoverability of $S\bar{X}, \ 
\bar{X}\sim\mathcal{N}\left(\mathbf{0},I_n\right)$. For reasons of ease of notation, we do not use 
this observation for the proof of Theorem~\ref{thm:gauss_privacy}.
\vspace{0.1cm}

The significance of Lemma~\ref{lem:A-sing} is that $\rho$-privacy under recoverability of $AX$ from $Z$ is equivalent to $\rho$-privacy under recoverability of $SV^TX$ from $\bar{Z}$. Recalling that $\tilde{S}$ 
is the $r\times n$-matrix consisting of the $r$ nonzero rows of $S$, the covariance matrix of the 
$\mathbb{R}^r$-valued rv $\tilde{S}V^TX$ is a diagonal matrix $\tilde{S}\tilde{S}^T$, an observation that facilitates establishing the converse for the proof of Theorem~\ref{thm:gauss_privacy}.

The following notation is relevant for the rest of the paper. For given $\mathbb{R}^m$-valued 
rvs $\Phi$ and $\phi$, let the components of $\Phi-\phi$ be denoted by 
$\left[\left(\Phi-\phi\right)_1,\ldots,\left(\Phi-\phi\right)_m\right]^T$. 
\begin{lemma}
\label{lem:dpi}
For $\rho\geq 0,$
\begin{equation}
\label{eq:add_rest}
    \sup_{P_{\bar{Z}|X} : \mathbb{E}
    \left[\left\Vert SV^{T}X-\bar{Z}\right\Vert^2\right]\leq\rho} \ \inf_{g} \ 
    \mathbb{E}\left[{\left\Vert X-g(\bar{Z})
\right\Vert}^2\right] = \sup_{\substack{P_{\tilde{Z}|X} : 
\mathbb{E}\left[\left\Vert \tilde{S}V^{T}X-\tilde{Z}\right\Vert^2\right]\leq\rho\\
\mathbb{E}\left[\left\lvert\left( \tilde{S}V^TX-\tilde{Z}\right)_i\right\rvert^2\right]
\leq s_i^2,\\i=1,\ldots,r}} \  \inf_{g} \ \mathbb{E}\left[{\left\Vert X-g(\tilde{Z})
\right\Vert}^2\right],
\end{equation} 
\noindent where the $\mathbb{R}^r$-valued rv $\tilde{Z}$ represents a generic stand-in for a $\rho$-QR under recoverability of $\tilde{S}V^{T}X$ and satisfies the additional constraints given under the 
supremum in the right-side of~\eqref{eq:add_rest}.
\end{lemma}
\vspace{0.2cm}

By Lemma~\ref{lem:dpi}, we can restrict the class of $\rho$-QRs $\bar{Z}$ for recovering $SV^{T}X$ in Lemma~\ref{lem:A-sing} to those specified $\tilde{Z}$ for the recoverability of $\tilde{S}V^TX$, as detailed under the 
supremum in the right-side of~\eqref{eq:add_rest} without any loss in 
$\rho$-privacy.\vspace{0.3cm}\\ 
\textit{Proof of Theorem~\ref{thm:gauss_privacy}}: Using 
Lemmas~\ref{lem:A-sing} and~\ref{lem:dpi}, we get in~\eqref{eq:gauss} that
\begin{align}
\pi(\rho)
&=\sup_{P_{Z|X} : \mathbb{E}\left[\left\Vert AX-Z\right\Vert^2\right]\leq\rho} \ 
\inf_g \ \mathbb{E}\left[{\left\Vert X-g(Z)
\right\Vert}^2\right]\nonumber\\
&=\sup_{P_{\bar{Z}|X} : \mathbb{E}\left[\left\Vert 
SV^{T}X-\bar{Z}\right\Vert^2\right]\leq\rho} 
\inf_g \ \mathbb{E}\left[{\left\Vert X-g(\bar{Z})
\right\Vert}^2\right]\nonumber\\
&=\sup_{\substack{P_{\tilde{Z}|X} : \mathbb{E}\left[\left\Vert \tilde{S}V^{T}X-
\tilde{Z}\right\Vert^2\right]\leq\rho\\\mathbb{E}\left[\left\lvert
\left( \tilde{S}V^TX-\tilde{Z}\right)_i\right\rvert^2\right]\leq s_i^2,\\i=1,\ldots,r}} \  
\inf_g \ \mathbb{E}\left[{\left\Vert X-g(\tilde{Z})
\right\Vert}^2\right].\label{eq:main_term}
\end{align}
\noindent We establish~\eqref{eq:thm_eqn}, with~\eqref{eq:main_term} serving as the springboard. First, we prove a converse proof showing that $\pi(\rho)$ cannot exceed the right-side of~\eqref{eq:main_term}. Then an achievability proof shows the reverse inequality by identifying an explicit $\rho$-QR that attains the right-side of~\eqref{eq:main_term} as $\rho$-privacy.

Starting with the converse, we have that for 
every $\mathbb{R}^r$-valued rv $\tilde{Z}$, generated according to a $P_{\tilde{Z}|X}$ 
satisfying the constraints in the right-side of~\eqref{eq:main_term},
\begin{equation}
\label{eq:constraints}
    \mathbb{E}\left[\left\Vert \tilde{S}V^{T}X-\tilde{Z}\right\Vert^2\right]\leq\rho, \ \ \ \ \ \ 
    \mathbb{E}\left[\left\lvert\left( \tilde{S}V^TX-\tilde{Z}\right)_i\right\rvert^2\right]\leq s_i^2,\ \ \ 
    i=1,\ldots,r, 
\end{equation}
so that
\begin{align}
&\inf_g \ \mathbb{E}\left[{\left\Vert X-g(\tilde{Z})
\right\Vert}^2\right]\nonumber\\
&\leq \mathbb{E}\left[{\left\Vert X-V\tilde{S}^T\left(\tilde{S}\tilde{S}^T\right)^{-1}\tilde{Z}
\right\Vert}^2\right]\label{eq:mmse_estim}\\
&=\mathbb{E}\left[{\left\Vert X-V\tilde{S}^T\left(\tilde{S}\tilde{S}^T\right)^{-1}
\tilde{S}V^TX+V\tilde{S}^T\left(\tilde{S}\tilde{S}^T\right)^{-1}
\tilde{S}V^TX-V\tilde{S}^T\left(\tilde{S}\tilde{S}^T\right)^{-1}\tilde{Z}
\right\Vert}^2\right]\nonumber\\
&=\mathbb{E}\left[{\left\Vert X-V\tilde{S}^T\left(\tilde{S}\tilde{S}^T\right)^{-1}\tilde{S}V^TX\right
\Vert}^2\right]\nonumber
\\&+\mathbb{E}\left[{\left\Vert V\tilde{S}^T\left(\tilde{S}\tilde{S}^T\right)^{-1}\tilde{S}V^TX-V\tilde{S}^T
\left(\tilde{S}\tilde{S}^T\right)^{-1}\tilde{Z}
\right\Vert}^2\right]\nonumber\\
&+2\mathbb{E}\left[X^T\left(I_n-V\tilde{S}^T\left(\tilde{S}\tilde{S}^T\right)^{-1}\tilde{S}V^T\right)^TV\tilde{S}^T
\left(\tilde{S}\tilde{S}^T\right)^{-1}\left(\tilde{S}V^TX-\tilde{Z}\right)\right]\nonumber\\
&=\text{mmse}\left(X|\tilde{S}V^TX\right)+
\mathbb{E}\left[{\left\Vert V\tilde{S}^T\left(\tilde{S}\tilde{S}^T\right)^{-1}
\left(\tilde{S}V^TX-\tilde{Z}\right)
\right\Vert}^2\right]\label{eq:mmse_XSVT}\\
&+2\mathbb{E}\left[X^T\left(\tilde{S}V^T-\tilde{S}V^TV\tilde{S}^T\left(\tilde{S}\tilde{S}^T\right)^{-1}\tilde{S}V^T\right)^T
\left(\tilde{S}\tilde{S}^T\right)^{-1}\left(\tilde{S}V^TX-\tilde{Z}\right)\right]\nonumber\\
&=\text{mmse}(X|U^TAX)+\mathbb{E}\left[{\left\Vert 
\tilde{S}^T\left(\tilde{S}\tilde{S}^T\right)^{-1}\left(\tilde{S}V^TX-\tilde{Z}\right)
\right\Vert}^2\right]\label{eq:12}\\
&=\text{mmse}(X|AX)+\sum_{i=1}^r\frac{1}{s_i^2}\mathbb{E}
\left[{\left\vert \left(\tilde{S}V^TX-\tilde{Z}\right)_i
\right\vert}^2\right]\label{eq:13}\\
&=n-r+\sum_{i=1}^r\frac{1}{s_i^2}\mathbb{E}\left[{\left\vert 
\left(\tilde{S}V^TX-\tilde{Z}\right)_i
\right\vert}^2\right],\label{eq:14}
\end{align}
where:~\eqref{eq:mmse_estim} uses the MMSE estimator of $X$ on the basis of $\tilde{S}V^TX$, leading 
to the first term in~\eqref{eq:mmse_XSVT}; the first term in~\eqref{eq:12} is obtained 
since $\tilde{S}$ is the matrix consisting of the nonzero rows of $S$ and~\eqref{eq:svd}; the 
second terms in~\eqref{eq:12} and~\eqref{eq:13}, and the first term in~\eqref{eq:14}, are 
obtained by the orthonormality of $V$,~\eqref{eq:S_matrix} 
and~\eqref{eq:mmse_XAX}, respectively. Considering the second term in the 
right-side of~\eqref{eq:14}, for $t=0,1,\ldots,r$, 
\begin{align}
 &  \sum_{i=1}^r\frac{1}{s_i^2}\mathbb{E}\left[{\left\vert 
\left(\tilde{S}V^TX-\tilde{Z}\right)_i
\right\vert}^2\right]\nonumber\\&=\sum_{i=1}^t\frac{1}{s_i^2}\mathbb{E}\left[{\left\vert 
\left(\tilde{S}V^TX-\tilde{Z}\right)_i
\right\vert}^2\right]+\sum_{i=t+1}^r\frac{1}{s_i^2}\mathbb{E}\left[{\left\vert 
\left(\tilde{S}V^TX-\tilde{Z}\right)_i
\right\vert}^2\right]\label{eq:first_term}\\
&\leq \sum_{i=1}^t\frac{1}{s_i^2}\mathbb{E}\left[{\left\vert 
\left(\tilde{S}V^TX-\tilde{Z}\right)_i
\right\vert}^2\right]+\frac{1}{s_{t+1}^2}\sum_{i=t+1}^r\mathbb{E}\left[{\left\vert 
\left(\tilde{S}V^TX-\tilde{Z}\right)_i
\right\vert}^2\right], \ \ \ \ \text{using~\eqref{eq:sing_val_asc}}\nonumber\\
&=\sum_{i=1}^t\left(\frac{1}{s_i^2}-\frac{1}{s_{t+1}^2}\right)\mathbb{E}\left[{\left\vert 
\left(\tilde{S}V^TX-\tilde{Z}\right)_i
\right\vert}^2\right]+\frac{1}{s_{t+1}^2}\sum_{i=1}^r\mathbb{E}\left[{\left\vert 
\left(\tilde{S}V^TX-\tilde{Z}\right)_i
\right\vert}^2\right]\nonumber\\
&\leq\sum_{i=1}^t\left(\frac{1}{s_i^2}-\frac{1}{s_{t+1}^2}\right)s_i^2+
\frac{\rho}{s_{t+1}^2}, \ \ \  \ \ \ \ \text{using~\eqref{eq:constraints}}\nonumber\\
&= t+\frac{\rho-\sum\limits_{i=1}^t s_i^2}{s_{t+1}^2},\label{eq:second_term}
\end{align}
\noindent from\footnotemark\footnotetext{In the right-side of~\eqref{eq:first_term}, the first 
and second terms are vacuous for $t=0$ and $t=r$, respectively. In the right-side 
of~\eqref{eq:second_term}, the second term and the summation in the second term are vacuous 
for $t=r$ and $t=0$, respectively.} which we get
\begin{equation}
\label{eq:111}
    \sum_{i=1}^r\frac{1}{s_i^2}\mathbb{E}\left[{\left\vert \left(\tilde{S}V^TX-\tilde{Z}\right)_i
\right\vert}^2\right]\leq \min\left\{\frac{\rho}{s_1^2},1+\frac{\rho-s_1^2}{s_2^2},
\ldots,r-1+\frac{\rho-\sum\limits_{i=1}^{r-1}s_i^2}{s_r^2},r\right\}.
\end{equation}
\noindent Using~\eqref{eq:111} in~\eqref{eq:14}, the right-side of~\eqref{eq:main_term}, 
and therefore $\pi(\rho)$, is bounded above as
\begin{equation}
    \label{eq:ub}
    \pi(\rho)\leq n-r+\min\left\{\frac{\rho}{s_1^2},1+\frac{\rho-s_1^2}{s_2^2},
    \ldots,r-1+\frac{\rho-\sum\limits_{i=1}^{r-1}s_i^2}{s_r^2},r\right\}.
\end{equation}

Next, we demonstrate achievability of the privacy in the right-side of~\eqref{eq:main_term} by 
describing explicitly a $\rho$-QR for the purpose. This $\rho$-QR represents an extension of the 
scheme in~\cite[Theorem $13$]{Wu12} in the separate context of maximizing (i.e., under worst-case 
noise) the MMSE of estimating a one-dimensional Gaussian rv on the basis of a one-dimensional noisy 
version of it under a constraint on the expected $\ell_2$-distance between the input and the noisy 
output. The scheme in~\cite[Theorem $13$]{Wu12} has the structure of attenuation of the input 
followed by additive independent Gaussian noise which will be the case for our achievability scheme, 
too, and is therefore an extension. To this end, by Lemmas~\ref{lem:A-sing},~\ref{lem:dpi}, it suffices 
to show a $\rho$-QR $\tilde{Z}=\tilde{Z}_o$ for the recoverability of $\tilde{S}V^TX$ and satisfies the 
constraints in~\eqref{eq:constraints}. Our $\rho$-QR is the $\mathbb{R}^r$-valued rv given by
\begin{equation}
\label{eq:gaussian_achiev}
\tilde{Z}_o=D_a\tilde{S}V^TX+D_{no}N
\end{equation}
\noindent where 
\begin{gather}
    D_a=\text{diag}\left(1-\frac{\rho_1}{s_1^2},\ldots,1-
    \frac{\rho_r}{s_r^2}\right),\nonumber\\
    D_{no}=\text{diag}\left(\sqrt{\rho_1-\frac{\rho^2_1}{s_1^2}},
    \ldots,\sqrt{\rho_r-\frac{\rho^2_r}{s_r^2}}\right)\label{eq:Diag_mat}
\end{gather}
\noindent where diag$\left(d_1,\ldots,d_r\right)$ denotes a diagonal matrix with (diagonal) elements $\left(d_1,\ldots,d_r\right)$, and \\$N\sim\mathcal{N}\left(\mathbf{0},I_r\right)$ is a  
$\mathbb{R}^r$-valued zero-mean Gaussian rv independent of $X$. Thus, $\tilde{Z}_o$ entails attenuating $\tilde{S}V^TX$ by $D_a$ and contaminating it with 
an additive independent Gaussian noise $D_{no}N$. 
The values of $\rho_1,\ldots,\rho_r$ in~\eqref{eq:Diag_mat} are chosen for various ranges of values of $\rho$ as follows.
\begin{align}
&\bullet \ 0\leq\rho \leq s_1^2: \ \ \    
\rho_1=\rho, \ \rho_2=\cdots=\rho_r=0;\nonumber\\
&\bullet \ s_1^2\leq\rho \leq s_1^2+s_2^2: \ \ \  \rho_1=s_1^2, \ \rho_2=\rho-s_1^2, \ 
\rho_3=\cdots=\rho_r=0;\nonumber\\
&\bullet \  s_1^2+s_2^2\leq\rho \leq s_1^2+s_2^2+s_3^2: \ \ \ \rho_1=s_1^2, \ \rho_2=s_2^2, \ 
\rho_3=\rho-s_1^2-s_2^2, \ \rho_4=\cdots=\rho_r=0;\nonumber\\
&\vdots\nonumber\\
& \bullet \   \sum\limits_{i=1}^{r-1}s_i^2\leq\rho\leq 
\sum\limits_{i=1}^r s_i^2: \ \ \  
\rho_1=s_1^2,\ldots,\rho_{r-1}=s_{r-1}^2, \ \rho_r=
\rho-\sum\limits_{i=1}^{r-1} s_i^2;\nonumber\\
&\bullet  \ \rho\geq \sum\limits_{i=1}^r s_i^2: \  \ \ 
\rho_1=s_1^2,\ldots,\rho_r=s_r^2.\nonumber\\\label{eq:rho_vals}
\end{align}
\noindent We show in Appendix~\ref{app:verification} that $\tilde{Z}_o$ as 
in~\eqref{eq:gaussian_achiev},~\eqref{eq:Diag_mat},~\eqref{eq:rho_vals} 
satisfies the constraints in~\eqref{eq:constraints} (with $\tilde{Z}=\tilde{Z}_o$). Observing that $X$ 
and $\tilde{Z}_o$ are jointly Gaussian, we have that the right-side of~\eqref{eq:main_term} is bounded below by 
\begin{align}
\inf_g \ \mathbb{E}\left[{\left\Vert X-g(\tilde{Z}_o)
\right\Vert}^2\right]&=\text{mmse}\left(X|\tilde{Z}_o\right)\nonumber\\
&=\mathbb{E}\left[\left\lVert X-\mathbb{E}\left[X\tilde{Z}_o^T\right]\left(\mathbb{E}\left[\tilde{Z}_o\tilde{Z}_o^T\right]\right)^{-1}\tilde{Z}_o\right\rVert^2\right]\nonumber
\\&=\mathbb{E}\left[\left( X-\mathbb{E}\left[X\tilde{Z}_o^T\right]\left(\mathbb{E}\left[\tilde{Z}_o\tilde{Z}_o^T\right]\right)^{-1}\tilde{Z}_o\right)^TX\right]\nonumber
\\
&=\text{tr}\left(\mathbb{E}\left[XX^T\right]\right)-\text{tr}\left(\mathbb{E}\left[
X\tilde{Z}_o^T\right]\left(\mathbb{E}\left[\tilde{Z}_o\tilde{Z}_o^T\right]\right)^{-1}\mathbb{E}\left[
\tilde{Z}_oX^T\right]\right)\nonumber\\
&=\text{tr}\left(I_n\right)-\text{tr}
\left(I_nV\tilde{S}^TD_a^T\left(D_a\tilde{S}V^TI_nV\tilde{S}^TD_a^T+
D_{no}I_rD_{no}^T\right)^{-1}D_a\tilde{S}V^TI_n\right)\nonumber\\
&=n-\text{tr}\left(\tilde{S}^TD_a^T
\left(D_a\tilde{S}\tilde{S}^TD_a^T+D_{no}D_{no}^T\right)^{-1}D_a\tilde{S}\right),\ \ \ \ \ \text{since $V$ is orthonormal}\label{eq:diag-comp}\\
&=n-r+\sum\limits_{i=1}^r
\frac{\rho_i}{s_i^2}\label{eq:mmse_XZo},
\end{align}
\noindent where the second term in the right-side of~\eqref{eq:diag-comp} is calculated using
\begin{align*}
D_a\tilde{S}\tilde{S}^TD_a^T&=\text{diag}\left(s_1^2+\frac{\rho_1^2}{s_1^2}-2\rho_1,\ldots,s_r^2+\frac{\rho_r^2}{s_r^2}-2\rho_r\right), \\
    D_{no}D_{no}^T&=\text{diag}\left(\rho_1-\frac{\rho_1^2}{s_1^2},\ldots,\rho_r-\frac{\rho_r^2}{s_r^2}\right),
      \end{align*}
\noindent and for $i\in\left\{1,\ldots,r\right\}, \ j\in\left\{1,\ldots,m\right\}$,
      \[\left(D_a\tilde{S}\right)_{ij}=\begin{cases}
      s_k-\frac{\rho_k}{s_k}, \ \ \ &i=j=k, \ k=1,\ldots,r\\
      0, \ \ \ &\text{otherwise},  
      \end{cases}
      \]
\noindent and
\[
      \tilde{S}^TD_a^T
\left(D_a\tilde{S}\tilde{S}^TD_a^T+D_{no}D_{no}^T\right)^{-1}D_a\tilde{S}=\text{diag}
\left(1-\frac{\rho_1}{s_1^2},\ldots,1-\frac{\rho_r}{s_r^2}\right).
\]
\noindent By recalling the equivalence of the right-sides of~\eqref{eq:thm_eqn} 
and~\eqref{eq:gauss-priv-alt-form} and substituting~\eqref{eq:rho_vals} in~\eqref{eq:mmse_XZo}, we get that the 
right-side of~\eqref{eq:main_term}, and therefore $\pi(\rho)$, is bounded below as
\begin{equation}
\label{eq:main_lb}
\pi(\rho)\geq n-r+\min\left\{\frac{\rho}{s_1^2},1+\frac{\rho-s_1^2}{s_2^2},
\ldots,r-1+\frac{\rho-\sum\limits_{i=1}^{r-1}s_i^2}{s_r^2},r\right\}.
\end{equation}

The theorem follows from~\eqref{eq:ub} and~\eqref{eq:main_lb}.
\qeed
\vspace{0.3cm}\\
\textit{Remark}: Recalling~\eqref{eq:svd}, let $\tilde{U}$ be the $m\times r$-matrix 
containing the first $r$ columns of $U$. The achievability scheme 
$\tilde{Z}_o=D_a\tilde{S}V^TX+D_{no}N$~\eqref{eq:gaussian_achiev}
is for maximizing privacy under recoverability of $\tilde{S}V^TX$. The corresponding 
achievability 
scheme or $\rho$-QR for the recoverability of $AX$, denoted by $Z_o$, is 
$Z_o=\tilde{U}D_a\tilde{U}^TAX+\tilde{U}D_{no}N$ 
which can be shown readily from the proof of Lemma~\ref{lem:A-sing}, and 
also has the same features of attenuation and independent additive Gaussian noise.

We conclude this section by extending $\rho$-privacy to the case when the querier 
wishes to compute an affine function $AX+b$ of the Gaussian user data 
$X\sim\mathcal{N}\left(\mathbf{0},I_n\right), \ n\geq 1$, for a given $A$ as in 
Section~\ref{sec:prelim}, and $b\in\mathbb{R}^m$. In Definition~\ref{def:rho-recov},~\eqref{eq:gauss} becomes
\begin{align}
\pi(\rho)&=\sup_{P_{Z|X} : \mathbb{E}\left[\left\Vert AX+b-Z\right
\Vert^2\right]\leq\rho}  \ \inf_{g} \ \mathbb{E}\left[{\left\Vert X-g\left(Z\right)
\right\Vert}^2\right]\label{gauss_affine}\\
&=\sup_{P_{Z|X} : \mathbb{E}\left[\left\Vert AX-(Z-b)
\right\Vert^2\right]\leq\rho}  \ \inf_{g} \ \mathbb{E}\left[{\left\Vert X-
g\left(Z-b\right)
\right\Vert}^2\right]\nonumber\\
&=\sup_{P_{\hat{Z}|X} : \mathbb{E}\left[\left\Vert AX-\hat{Z}\right\Vert^2
\right]\leq\rho}  \ \inf_{g} \ \mathbb{E}\left[{\left\Vert X-g\left(\hat{Z}\right)
\right\Vert}^2\right]\nonumber
\end{align}
from which we conclude that $\pi(\rho)$ as in~\eqref{gauss_affine} is equal 
to the right-side of~\eqref{eq:thm_eqn}. Observe that $\pi(\rho)$ does not 
depend on $b$, as is to be expected.

\section{Discussion}
\label{sec:discussion}

We give a heuristic explanation of the form of $\pi(\rho)$ in~\eqref{eq:thm_eqn}. By Lemma~\ref{lem:A-sing}, note that the recoverability of $AX$ is equivalent to the recoverability of $SV^TX$, and therefore $\tilde{S}V^TX$, which consists of $r$ components. We recall that $\tilde{S}$ is the matrix composed of the $r$ nonzero rows of $S$, due to which $AX$ consists effectively of $r$ components corresponding to the $r$ singular values of $A$. At $\rho=0$, the querier is provided the exact value of $AX$. From~\eqref{eq:gaussian_achiev},~\eqref{eq:Diag_mat},~\eqref{eq:rho_vals}, observe that as $\rho$ increases to $s_1^2$, the component of $AX$ corresponding to the smallest singular value of $A$ is concealed from the querier. As $\rho$ increases further, more components of $AX$ are hidden from the querier, with each subsequent component corresponding to a larger singular value of $A$. This explains the piecewise affine form of $\pi(\rho), \ \rho\geq 0$, in~\eqref{eq:thm_eqn}. Therefore, for lower values of $\rho$, i.e., in the high recoverability regime, only those components that correspond to smaller singular values of $A$, can be concealed from the querier.

This work is an initial foray into tackling the larger objective of characterizing data privacy under function recoverability, where the
data is of the analog type. Therefore, several open questions remain some of which are stated next.

For reasons of mathematical tractability, we have assumed that the covariance 
matrix of the Gaussian user data $X$ is $I_n$. The problem of computing $\rho$-privacy 
for an arbitrary (positive-definite) covariance is open. We conjecture 
that the form of $\pi(\rho)$ (piecewise affine in $\rho$) and the structure of the 
achievability scheme (attenuation and independent additive Gaussian noise) in Theorem~\ref{thm:gauss_privacy}, will hold.

A natural extension of this work involves the querier obtaining from the 
user multiple QRs each satisfying the $\rho$-recoverabilty condition. Specifically, 
given user data $X\sim\mathcal{N}\left(\mathbf{0},I_n\right)$, a querier receives 
multiple $\rho$-QRs $Z_1,\ldots,Z_t, \ t\geq 1,$
each satisfying~\eqref{eq:recov1} and generated by conditional distributions 
$P_{Z_1|X},\ldots,P_{Z_t|X}$. The $\rho$-QRs are taken
to be conditionally mutually independent, conditioned on $X$,
but not necessarily identically distributed. Correspondingly, for each $\rho\geq 0$ 
and $t\geq 1,$ the $\rho$-privacy $\pi_t(\rho)$ is defined as
\[
\pi_t(\rho)=\sup_{\substack{P_{Z_1|X},\ldots,P_{Z_t|X}:\\\mathbb{E} 
\left[\left\lVert AX-Z_i\right\rVert^2\right]\leq \rho,  \ i=1,\ldots,t}} \ 
\inf_{g_t} \ \mathbb{E} \left[\left\lVert X-g_t\left(Z_1,\ldots,Z_t\right)
\right\rVert^2\right]
\]

\noindent where the infimum is taken over all estimators 
$g_t:\mathbb{R}^{m\times  t}\rightarrow\mathbb{R}^n$ of $X$ on the basis 
of $Z_1,\ldots,Z_t$. The task is to characterize $\pi_t(\rho)$ and obtain 
the rate of decay of $\pi_t(\rho)$ with $t$. A candidate for the for $\rho$-QRs is~\eqref{eq:gaussian_achiev} with mutually 
independent Gaussian noise rvs added to them. Will this be optimal in attaining $\pi_t(\rho)$? 

Another broader extension of this work entails recoverability and privacy 
being measured by $\ell_p$-distance and $\ell_q$-distance criteria, 
$p,q\geq 1$, respectively. We seek a characterization of $\pi_{p,q}(\rho)$ given by
\[\pi_{p,q}(\rho)=
\sup_{P_{Z|X} : \mathbb{E}\left[\left\Vert AX-Z\right\Vert^p\right]\leq\rho} \ 
\inf_{g} \ \mathbb{E}\left[{\left\Vert X-g(Z)
\right\Vert}^q\right].
\]
\noindent Does the structure of the solution in Theorem~\ref{thm:gauss_privacy} 
change?

Finally, it is also of interest to examine $\rho$-privacy under recoverability of 
an arbitrary but given measurable function $f:\mathbb{R}^n\rightarrow\mathbb{R}^m, \ 
m\geq 1,$ not limited to being linear. This problem, of a more demanding nature due to 
the (possible) nonlinearity of the mapping $f$, requires a new approach.



\appendices
\section{Calculation of $\text{var}(AX)$ and $\text{mmse}\left(X|AX\right)$}
\label{app:gauss_standard_calc}
\noindent We have 
\[
\text{var}(AX)=\text{tr}\left(AA^T\right)=\text{tr}\left(USV^TVS^TU^T\right)=
\text{tr}\left(USS^TU^T\right)=\text{tr}\left(SS^T\right)=\sum_{i=1}^rs_i^2.
\]

\noindent Next, noting that $X$ and $AX=USV^TX$ are jointly Gaussian, we have
\begin{align*}
&\text{mmse}\left(X|AX\right)=\text{mmse}\left(X|USV^TX\right)
\\&=\mathbb{E}\left[\left\lVert X-\mathbb{E}\left[X\left(USV^TX\right)^T\right]\left(\mathbb{E}\left[USV^TX\left(USV^TX\right)^T\right]\right)^{-1}USV^TX\right\rVert^2\right]
\\&=\mathbb{E}\left[\left( X-\mathbb{E}\left[X\left(USV^TX\right)^T\right]\left(\mathbb{E}\left[USV^TX\left(USV^TX\right)^T\right]\right)^{-1}USV^TX\right)^TX\right]
\\&=\text{tr}\left(\mathbb{E}\left[XX^T\right]\right)\\&\hspace{2cm}-\text{tr}\left(\mathbb{E}\left[
X\left(USV^TX\right)^T\right]\left(\mathbb{E}\left[\left(USV^TX\right)\left(USV^TX\right)^T\right]\right)^{-1}\mathbb{E}\left[
USV^TXX^T\right]\right)\nonumber\\
&=\text{tr}\left(I_n\right)-\text{tr}
\left(I_nVS^TU^T\left(USV^TI_nVS^TU^T\right)^{-1}USV^TI_n\right)\nonumber\\
&=n-\text{tr}\left(S^T\left(SS^T\right)^{-1}S\right)=n-r.
\end{align*}

\section{Proofs of Lemmas~\ref{lem:A-sing},~\ref{lem:dpi}}
\label{app:gauss_priv}

\noindent\textbf{Proof of Lemma~\ref{lem:A-sing}}: Recalling~\eqref{eq:svd}, we have
\[
    \mathbb{E}\left[\left\Vert AX-Z\right\Vert^2\right]=\mathbb{E}\left[\left\Vert 
    U\left(U^TAX-U^TZ\right)\right\Vert^2\right]=\mathbb{E}\left[\left\Vert SV^TX-U^TZ\right\Vert^2\right],
\]
from which we get
\begin{align*}
&    \sup_{P_{Z|X} : \mathbb{E}\left[\left\Vert AX-Z\right\Vert^2\right]\leq\rho} \ 
    \inf_{g} \ \mathbb{E}\left[{\left\Vert X-g(Z)
\right\Vert}^2\right]\\&= \sup_{P_{Z|X} : 
\mathbb{E}\left[\left\Vert SV^{T}X-U^TZ\right\Vert^2\right]\leq\rho} \inf_{g} \ \mathbb{E}\left[{\left\Vert X-g(U^TZ)
\right\Vert}^2\right]\\
&= \sup_{P_{\bar{Z}|X} : 
\mathbb{E}\left[\left\Vert SV^{T}X-\bar{Z}\right\Vert^2\right]\leq\rho} \inf_{g} \ 
\mathbb{E}\left[{\left\Vert X-g(\bar{Z})
\right\Vert}^2\right], \ \ \ \ \text{since $U$ is orthonormal}.
\end{align*}
 \qeed

\vspace{0.3cm}

\noindent\textbf{Proof of Lemma~\ref{lem:dpi}}: Since the supremum in the 
right-side of~\eqref{eq:add_rest} is over a restricted
set compared with the left-side, it suffices to show 
that~\eqref{eq:add_rest} holds with 
``$\leq,$'' i.e.,
\begin{equation}
    \label{eq:gauss-priv-eqn-toshow}
     \sup_{P_{\bar{Z}|X} : \mathbb{E}
    \left[\left\Vert SV^{T}X-\bar{Z}\right\Vert^2\right]\leq\rho} \ \inf_{g} \ 
    \mathbb{E}\left[{\left\Vert X-g(\bar{Z})
\right\Vert}^2\right] \leq \sup_{\substack{P_{\tilde{Z}|X} : 
\mathbb{E}\left[\left\Vert \tilde{S}V^{T}X-\tilde{Z}\right\Vert^2\right]\leq\rho\\
\mathbb{E}\left[\left\lvert\left( \tilde{S}V^TX-\tilde{Z}\right)_i\right\rvert^2\right]
\leq s_i^2,\\i=1,\ldots,r}} \  \inf_{g} \ \mathbb{E}\left[{\left\Vert X-g(\tilde{Z})
\right\Vert}^2\right].
\end{equation}

Since $S$ contains only $r$ nonzero rows, $SV^{T}X$ in the left-side of~\eqref{eq:add_rest} is a $\mathbb{R}^m$-valued rv containing at most $r$ nonzero elements. Recalling that $\tilde{S}$ is the 
$r\times n$-matrix consisting of the nonzero rows of $S$, it is easily seen that
\begin{equation}
\label{eq:m-to-r}
     \sup_{P_{\bar{Z}|X} : \mathbb{E}
    \left[\left\Vert SV^{T}X-\bar{Z}\right\Vert^2\right]\leq\rho} \ \inf_{g} \ 
    \mathbb{E}\left[{\left\Vert X-g(\bar{Z})
\right\Vert}^2\right]= \sup_{P_{\check{Z}|X} : \mathbb{E}
    \left[\left\Vert \tilde{S}V^{T}X-\check{Z}\right\Vert^2\right]\leq\rho} \ \inf_{g} \ 
    \mathbb{E}\left[{\left\Vert X-g(\check{Z})
\right\Vert}^2\right]
\end{equation}
where $\check{Z}$ is a $\mathbb{R}^r$-valued rv denoting a $\rho$-QR under recoverability of $\tilde{S}V^{T}X$. 
For every $\mathbb{R}^r$-valued rv $\check{Z}=\left[\check{Z}_1,\ldots,\check{Z}_r\right]^T$, 
consider the $\mathbb{R}^r$-valued rv $\tilde{Z}$ given by 
\begin{equation}
\label{eq:derived-rv}
    \tilde{Z}=\left[\check{Z}_1\mathbbm{1}\left(\mathbb{E}\left[\left\lvert\left( \tilde{S}V^TX-\check{Z}\right)_1\right\rvert^2\right]
\leq s_1^2\right),\ldots,\check{Z}_r\mathbbm{1}\left(\mathbb{E}\left[\left\lvert\left( \tilde{S}V^TX-\check{Z}\right)_r\right\rvert^2\right]
\leq s_r^2\right)\right]^T.
\end{equation}
\noindent Observe that
\begin{align}
\mathbb{E}
    \left[\left\vert \left(\tilde{S}V^{T}X-\tilde{Z}\right)_i\right\vert^2\right]&=\begin{cases}
    \mathbb{E}\left[\left\vert \left(\tilde{S}V^{T}X-\check{Z}\right)_i\right\vert^2\right], \ \ &\mathbb{E}\left[\left\vert \left(\tilde{S}V^{T}X-\check{Z}\right)_i\right\vert^2\right]\leq s_i^2\\
    s_i^2, \ \ &\mathbb{E}\left[\left\vert \left(\tilde{S}V^{T}X-\check{Z}\right)_i\right\vert^2\right]>s_i^2
    \end{cases}\label{eq:app-addn-const}\\&\leq  \mathbb{E}\left[\left\vert \left(\tilde{S}V^{T}X-\check{Z}\right)_i\right\vert^2\right], \ \ \ i=1,\ldots,r, \nonumber
\end{align}
from which, owing to the constraint under the supremum in the right-side of~\eqref{eq:m-to-r}, we get
\begin{equation}
\label{eq:inequality}
\mathbb{E}
    \left[\left\Vert \tilde{S}V^{T}X-\tilde{Z}\right\Vert^2\right]\leq \mathbb{E}
    \left[\left\Vert \tilde{S}V^{T}X-\check{Z}\right\Vert^2\right]\leq\rho.
\end{equation}
Since
\begin{equation*}
    X\MC\check{Z}\MC \tilde{Z},
\end{equation*}
and using data processing 
inequality for MMSE~\cite{Zamir98},~\cite{Wu12}, we get
\begin{equation}
    \label{eq:MC}
  \inf_{g} \ 
    \mathbb{E}\left[{\left\Vert X-g(\check{Z})
\right\Vert}^2\right]\leq  \inf_{g} \ \mathbb{E}\left[{\left\Vert X-g(\tilde{Z})
\right\Vert}^2\right].
\end{equation}
\noindent We have shown that for every $\mathbb{R}^r$-valued rv $\check{Z}$ that satisfies the $\rho$-recoverability constraint, there exists another $\mathbb{R}^r$-valued rv $\tilde{Z}$~\eqref{eq:derived-rv} that also satisfies the $\rho$-recoverability constraint~\eqref{eq:inequality} and additionally, due to~\eqref{eq:app-addn-const}, meets the constraints
\begin{equation}
    \label{eq:add-const-formal}
    \mathbb{E}\left[\left\lvert\left( \tilde{S}V^TX-\tilde{Z}\right)_i\right\rvert^2\right]
\leq s_i^2,\ \ \ \ i=1,\ldots,r.
\end{equation}
Therefore, using~\eqref{eq:inequality},~\eqref{eq:MC},~\eqref{eq:add-const-formal}, we get
\[
\sup_{P_{\check{Z}|X} : \mathbb{E}
    \left[\left\Vert \tilde{S}V^{T}X-\check{Z}\right\Vert^2\right]\leq\rho} \ \inf_{g} \ 
    \mathbb{E}\left[{\left\Vert X-g(\check{Z})
\right\Vert}^2\right]\leq\sup_{\substack{P_{\tilde{Z}|X} : 
\mathbb{E}\left[\left\Vert \tilde{S}V^{T}X-\tilde{Z}\right\Vert^2\right]\leq\rho\\
\mathbb{E}\left[\left\lvert\left( \tilde{S}V^TX-\tilde{Z}\right)_i\right\rvert^2\right]
\leq s_i^2,\\i=1,\ldots,r}} \  \inf_{g} \ \mathbb{E}\left[{\left\Vert X-g(\tilde{Z})
\right\Vert}^2\right],
\] 
\noindent which along with~\eqref{eq:m-to-r} gives~\eqref{eq:gauss-priv-eqn-toshow}.
\qeed

\section{Verification that $\tilde{Z}_o$ satisfies~\eqref{eq:constraints}}

\label{app:verification}

\noindent For $i=1,\ldots,r$,
\begin{align}
     \mathbb{E}\left[\left\lvert\left( \tilde{S}V^TX-\tilde{Z}_o\right)_i\right\rvert^2\right]&=\mathbb{E}\left[\left\lvert\left( \tilde{S}V^TX-D_a\tilde{S}V^{T}X-D_{no}N\right)_i\right\rvert^2\right]\nonumber\\
     &=\mathbb{E}\left[\left\lvert\left( \tilde{S}V^TX-D_a\tilde{S}V^{T}X\right)_i\right\rvert^2\right]+\rho_i-\frac{\rho_i^2}{s_i^2}\nonumber\\
          &=\frac{\rho_i^2}{s_i^2}+\rho_i-\frac{\rho_i^2}{s_i^2}=\rho_i\label{eq:app-gauss-eqn}\\
          &\leq s_i^2.\nonumber
\end{align}
\noindent Next,
\begin{align*}\mathbb{E}\left[\left\Vert \tilde{S}V^{T}X-\tilde{Z}_o\right\Vert^2\right]&= \sum_{i=1}^r\mathbb{E}\left[\left\lvert\left( \tilde{S}V^TX-D_a\tilde{S}V^{T}X-D_{no}N\right)_i\right\rvert^2\right]\\
    &=\sum_{i=1}^r \rho_i, \ \  \ \ \text{using~\eqref{eq:app-gauss-eqn}}\\
    &=\rho.
\end{align*}

\section*{Acknowledgments}

The author thanks Prakash Narayan for
the many helpful discussions of this work from~\cite{Nages23} and detailed comments
on this manuscript. The author also thanks Shun Watanabe for useful discussions and an 
anonymous referee~\cite{Nages22-2} for the observation contained in the Remark after Lemma~\ref{lem:A-sing}.

\end{document}